\documentclass[pra,twocolumn]{revtex4}
\usepackage{amssymb,amsfonts,amsmath,graphicx}
\begin{document}

\title{Counterfactual reasoning in time-symmetric quantum mechanics}

\author{D. J. Miller}

\affiliation{Centre for Time, Department of Philosophy, Main Quad A14, University of Sydney NSW 2006, Australia}
\email{D.Miller@unsw.edu.au}
\altaffiliation[Permanent Address: ]{School of Physics, The University of New South Wales, Sydney NSW 2052, Australia}
\begin{abstract}
A qualification is suggested for the counterfactual reasoning involved in some aspects of time-symmetric quantum theory (which involves ensembles selected by both the initial and final states).  The qualification is that the counterfactual reasoning should only apply to times when the quantum system has been subjected to physical interactions which place it in a ``measurement-ready condition" for the unperformed experiment on which the counterfactual reasoning is based.  The defining characteristic of a ``measurement-ready condition" is that a quantum system could be found to have the counterfactually ascribed property without direct physical interaction with the eigenstate corresponding to that property.
\end{abstract}

\maketitle

\section{Introduction}

While standard quantum mechanics (SQM) involves ensembles defined or preselected by the preparation state, time-symmetric quantum theory (TSQT) \cite{avtsqtreview}, also referred to as the two-state vector formalism, involves ensembles both preselected by the preparation state and postselected by considering only those cases which give a chosen measurement outcome. In the context of TSQT, Aharonov, Vaidman and co-authors \cite{aaaprl,vaaprl,avjpa,vprl} have drawn a number of interesting conclusions based on counterfactual reasoning about different {\it gedanken} experiments. From the beginning \cite{bub,sharp1}, concerns have been expressed about the legitimacy of the counterfactual reasoning involved in some of the conclusions in TSQT \cite{sharp2,cohen,cohenhiley,blasi,miller,kast1,kast2,kast3,kast4,kast5}.  Initially, there was a misunderstanding on the part of some authors (for example, \cite{miller}) that assertions about ``elements of reality" on the basis of the counterfactual reasoning were being made in an ontological sense. It has now been made clear that that is not the case and the counterfactual reasoning in TSQT has been defended \cite{vcohen,vff29,vaidmancfs,vaidman1,vaidman4}. Recently a new example reminiscent of the much-discussed 3-box example of time-symmetric counterfactual reasoning \cite{avtsqtreview,aaaprl,avjpa,bub,sharp1,sharp2,vff29,vaidmancfs,kast2,kast3,kast4,kast5,kentprl,griffithspra,marchildon,kirkpatrick} has been proposed \cite{avpranslits} and discussed \cite{kastnslits}.

The counterfactual reasoning involved in TSQT relies on the probabilities of SQM for the outcomes of successive measurements of a quantum system. The probabilities were first considered by Aharonov, Bergmann and Lebowitz \cite{abl} and the expression for the probabilities is often referred to as the ABL rule. Kastner has emphasised the significance for counterfactual reasoning in TSQT of specifying the observables that are considered to be actually measured in the application of the ABL rule and has concluded that the ABL rule cannot be used in a counterfactual sense except in cases when the criteria for consistent histories are satisfied {\cite{kast2,kast3,kast4,kast5}.

In the current work the significance of the observable involved in the ABL rule is taken into account in a specific way but it is argued that the counterfactual reasoning can be retained in all cases if it is  subject to a qualification. Some consequences of the suggested qualification are then explored. 

\section{Counterfactual reasoning}

The counterfactual reasoning in TSQT can be dealt with by considering two identical quantum systems, each in its own copy of a complex Hilbert space $\mathcal{H}$. One quantum system is in the real world (RW) in which we live and the other quantum system is in a counterfactual world (CFW). For both quantum systems, the initial state at time $t_a$ is the ray $| a \rangle$, an eigenstate of an observable $A$.  (Throughout the presentation, all states will be represented by rays in $\mathcal{H}$ because nothing turns on making that simplification.)  For the quantum system in the CFW a measurement of observable $C$ is made at time $t_c > t_a$, and the counterfactual quantum system is found to be in one of the non-degenerate eigenstates $|c_i \rangle $ with corresponding eigenvalue $c_i$. For the quantum system in the RW no measurement of $C$ is made.  For both quantum systems, a measurement of observable $B$ is made at time $t_b > t_c$ and only those cases are considered for which the final state at time $t_b$ is the ray $| b \rangle$.  That is, an ensemble in the RW and an ensemble in the CFW are selected on the basis that the initial state and the final states are $| a \rangle$ and $| b \rangle$ respectively. 

The counterfactual reasoning involves attributing (unmeasured) properties $c_i$ to the quantum system in the preselected and postselected RW ensemble with the same probabilities as the probabilities for the outcomes $c_i$ of measurements performed on the preselected and postselected CFW ensemble. In the case that the probability for an outcome $c_i$ in the CFW world is unity, $c_i$ is said \cite{vff29,vaidmancfs} to be an ``element of reality" (in a non-ontological sense) in the RW.

It is apparent \cite{sharp1} that the ensembles selected by imposing the above selection criterion in the RW and the CFW are different conceptually and are usually also different numerically. They are different conceptually because the quantum system in the CFW is subject to an intermediate measurement at time $t$ and the quantum system in the RW is not. They are different numerically when a different proportion of all possible cases is selected into the ensemble as a consequence of the measurement carried out at $t_c$ in the CFW compared with the measurement not being carried out in the RW.  At the extremes, if $\langle a | c_i \rangle = 0$ and/or $\langle c_i | b \rangle = 0$ then the preselected and postselected ensemble will be empty in the CFW in which $c_i$ is the outcome at $t_c$, although the ensemble will not be empty in the RW (unless $\langle a | b \rangle = 0$).  If $\langle a | b \rangle = 0$, the ensemble will be empty in the RW, although it will not be in the CFW provided $\langle a | c_i \rangle \neq 0$ and $\langle c_i | b \rangle \neq 0$. Whether or not the differences between the ensembles are significant has been the subject of discussion \cite{bub,sharp1,sharp2,cohen,cohenhiley,blasi,miller,vcohen,vff29,vaidmancfs,kast1,kast2,kast3,kast4,kast5,vaidman1,vaidman4}. For the purposes of the present work, it will be assumed that the differences between the ensembles are not significant. 

Aharonov, Bergmann and Lebowitz (ABL) \cite{abl} have shown how SQM can be used to calculate the probabilities of the sequence of measured properties $a \rightarrow c_i \rightarrow b$ in what we have defined as the CFW.  The conditional probability that the outcome of the measurement of $C$ in CFW is $c_i$ at time $t_c$, with $t_a < t_c < t_b$, given the initial state $a$ at $t_a$ and final state $b$ at $t_b$ is
\begin{equation}
\text{Pr}^{\text{ABL}}[c_i|a,C,b] = \frac{|\langle a| c_i \rangle \langle c_i| b \rangle|^2}{\sum_j |\langle a| c_j \rangle \langle c_j| b \rangle|^2}.
\end{equation}
The notation suggested by Kastner \cite{kast3} has been adopted in Eq.~(1) (in a slightly modified form) to indicate expressly that a measurement of the observable $C$ has been made in the period between $t_a$ and $t_b$.

Because the ABL rule involves a sequence of measured properties, the quantum system in each measured state must be left in an accessible form after the measurement (except perhaps the final measurement) in order that the subsequent measurements can take place.  That condition is satisfied by measurements of the first kind \cite{piron}.  An essential point is that, in order for example to carry out the measurement of $C$ and follow it by a measurement of $B$,  the quantum system must be subjected to a physical interaction which allows the eigenstates of $C$, other than the desired state $c_i$, to be filtered out. A possible sequence is shown schematically in Fig.~1(a) where the quantum system is prepared at time $t_a$ in the state $|a \rangle $, is subjected to a physical interaction at time $t_{c1}$ which allows the subsequent measurement of $C$ to take place at time $t_c$ and is then subjected to a further physical interaction at time $t_{c2}$ which allows the final measurement of observable $B$ to take place at time $t_b$.

\begin{figure}
\includegraphics[scale=0.4]{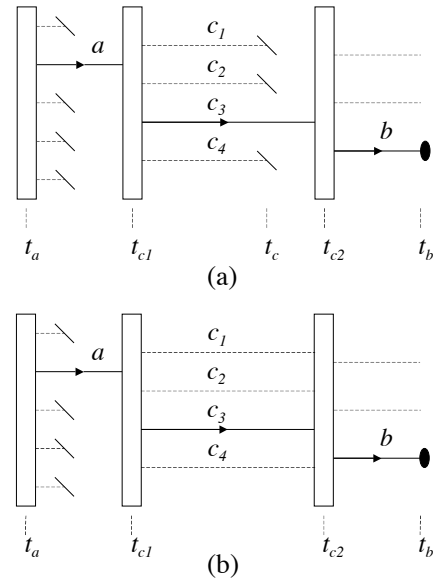}
\caption{A possible sequence of physical interactions that could be used for counterfactual reasoning about eigenstate $|c_3 \rangle $ of observable $C$, given preselection for state $|a \rangle$  and postselection for state $|b \rangle$. (a) In the counterfactual world (CFW), an experiment is actually performed with the outcome $|c_3 \rangle $. (b) In the real world (RW), no experiment is performed but it is here argued that counterfactual reasoning requires that the quantum system be subject to the necessary preliminary physical interactions which put it into a ``measurement-ready condition".}
\label{fig1}
\end{figure}

The sequence shown in Fig.~1(a) for performing the measurement of $C$ gives the same probability as any other method of measuring $C$. The advantages of the method shown in Fig.~1(a) is that (i) it allows (as required) the subsequent measurement of observable $B$ to take place and (ii) once the physical interaction at $t_{c1}$ has taken place the measurement outcome $c_i$ involved in Eq.~(2) can be obtained by blocking the states with $j \neq i$ and retaining only the cases when a null result at each block is obtained. Note that in the latter step, there is no direct interaction with the state $c_i$. Specifically, the Hamiltonian $H$ representing the measurement of the states $c_j$, $j \neq i$ has no matrix elements with $c_i$, i.e. $\langle c_j|H|c_i \rangle = 0$ for all $j$.

For $t_{c1} < t < t_{c}$ the quantum system can be said to be in a ``measurement-ready condition" for observable $C$. Placing the quantum system in a measurement-ready condition involves a unitary time evolution. If the measurement at $t_c$ is not carried out the  measurement-ready condition is reversible, for example by a choosing the interaction at $t_{c2}$ in Fig.~1(b) as the inverse of the interaction at $t_{c1}$. On the other hand, the measurement of $c_i$ which takes place in the CFW, involving the actual blocking of paths for $j \neq i$, is a non-unitary process and is irreversible. The concept of a ``partial measurement" in the 3-box problem \cite{kast5} seems to lie between these two extremes.

In the way the counterfactual reasoning in TSQT has been used to date \cite{avtsqtreview,aaaprl,vaaprl,avjpa,vprl,vcohen,vff29,vaidmancfs,vaidman1,vaidman4}, the probability in Eq. (1) is assigned to the quantum system in the RW for the outcome $c_i$ of the (unperformed) measurement of property $C$ for the whole of the period $t_a < t < t_b$. That is, it has been assumed that
\begin{equation}
\text{Pr}^{\text{RW}}[c_i|a,-,b] = \text{Pr}^{\text{ABL}}[c_i|a,C,b] \;\;\; t_a < t < t_b.
\end{equation}
where $\text{Pr}^{\text{RW}}[c_i|a,-,b]$ indicates that the quantum system in the RW has neither been measured for observable $C$ (unlike the quantum system in the CFW) nor placed in a measurement-ready condition for the measurement required for the ABL rule to apply in the CFW.

It is the purpose of the present work to suggest that the relationship in Eq. (2) requires a qualification. The qualification is that no inferences about the quantum system in the RW can be drawn from experiments performed on the quantum system in the CFW unless the quantum system in the RW is subjected to a sequence of physical interactions that places it in a measurement-ready condition for the ABL measurements required to be carried out on the quantum system in the CFW. That is, in order for the counterfactual reasoning to apply in the RW, the quantum system in the RW should be as shown in Fig. 1(b) where the quantum system can be seen to be in a measurement-ready condition for observable $C$ in the period $t_{c1} < t < t_{c2}$. In those cases only, a property like $c_i = c_3$ in the case shown in Fig. 1(b) can be ascribed to the quantum system in the RW and then only for the period $t_{c1} < t < t_{c2}$.

The following definition is proposed:\\[0.1in]
\noindent
\underline{\it Definition A}
\noindent
Time-symmetric counterfactual reasoning in relation to observable $C$ is valid only in the time period $t_{c1} < t < t_{c2}$ when the quantum system in the RW is in a measurement-ready condition for observable $C$. A quantum system is in a ``measurement-ready condition" for observable $C$ when it could be ascertained whether or not the quantum system has the property corresponding to any one of the eigenstates $c_i$ of $C$ by performing a measurement described by a Hamiltonian which does not couple with $c_i$.\\

Thus the counterfactual probability that an (unperformed) measurement on the quantum system in the RW would yield the outcome $c_i$ is 
\begin{equation}
\text{Pr}^{\text{RW}}[c_i|a,(C),b] = \left\{ \begin{array}{l}
\text{Pr}^{\text{ABL}}[c_i|a,C,b] \;\; t_{c1} < t < t_{c2} \\
\text{undefined otherwise}  
\end{array}
\right.
\end{equation}
where $t_{c1}$ and $t_{c2}$ are the beginning and end of the period sometime during which the measurement is carried out in the CFW. The notation $(C)$ in Prob$[c_i|a,(C),b]$ means that the quantum system is ready for a measurement of $C$ to be performed for $t_{c1} < t < t_{c2}$ but the measurement is not actually carried out in the RW. The measurement is carried out in the CFW where, as a consequence, the ABL rule can be applied.

There is a corresponding modification to the definition of (non-ontological) elements of reality given previously by Vaidman \cite{vaidmancfs} as follows:\\[0.1in]
\noindent
\underline{\it Definition B}
If the probability of the result $C = c_i$ of a measurement of observable $C$ at time $t$ in the CFW is unity, then there exists an element of reality $C=c_i$ for a quantum system in the RW provided the quantum system in RW is in a measurement-ready condition for $C$ at time $t$.\\

If the above conditions on counterfactual reasoning are accepted, then some of the conclusions that have been drawn from counterfactual reasoning in TSQT need to be modified.  The types of modifications are next identified by considering several of the well-known examples.

\section{Applications}

\subsection{Dispersion-free values of non-commuting observables}

It follows from the ABL rule in Eq.~(1) that $\text{Pr}^{\text{ABL}}[a|a,A,b] = 1$ and  $\text{Pr}^{\text{ABL}}[b|a,B,b] = 1$. From counterfactual reasoning based on Eq. (2), that is without confining the counterfactual reasoning about observable $A$ (and similarly for $B$) to the period when the quantum system is subject to the physical interaction required to be measurement-ready for the measurement of $A$ (and similarly for $B$), it has been concluded \cite{avtsqtreview,aaaprl,vaaprl,avjpa,vprl,vcohen,vff29,vaidmancfs,vaidman1,vaidman4} that since $\text{Pr}^{\text{RW}}[a|a,-,b] = 1$ and $\text{Pr}^{\text{RW}}[b|a,-,b] = 1$, then the quantum system in both the RW and the CFW must have (in a non-ontological sense) definite, dispersion-free values of both $A$ and $B$ (namely $a$ and $b$ respectively) throughout the period $t_a < t < t_b$.

That result does not follow if Eq. (3) instead of Eq. (2) is used for the counterfactual reasoning. The reasoning can be based on Fig. 1 by eliminating the physical transformation for the intervening measurement of observable $C$, i.e. by making $t_{c1} = t_{c2}$ in Fig. 1.  From Eq. (3), which is the present suggestion, the counterfactual reasoning should be based on the following results
\begin{subequations}
\begin{eqnarray}
\text{Pr}^{\text{RW}}[a_i|a,(A),b] & = & \left\{ \begin{array}{ll}
1 & t_a < t < t_{c2} \\
\text{undefined} &  t > t_{c2}  
\end{array}
\right. \\
\text{Pr}^{\text{RW}}[b_i|a,(B),b] & = & \left\{ \begin{array}{ll}
\text{undefined} & t < t_{c2}  \\
1 & t_{c2} < t < t_b    
\end{array}
\right. .
\end{eqnarray}
\end{subequations}

Therefore the only conclusion that can be drawn is that property $A=a$ is an element of reality for $ t_a < t < t_{c2}$ and property $B=b$ is an element of reality for $t_{c2} < t < t_b$. Consequently, it cannot be said that the quantum system has definite, dispersion-free values of the non-commuting observables $A$ and $B$ at any single time. Non-commuting operators could never simultaneously be assigned definite, dispersion-free values on the present approach because it is not possible to place the quantum system in the RW (nor in the CFW) in a physical situation where the appropriate measurements could be carried out at the same time.

One of the specific conclusions on the basis of the previous counterfactual reasoning has been that two non-commuting components of the spin of a quantum system can be ascertained at a given time on the basis of preselection for one spin component and postselection for the other \cite{vaaprl,avjpa,vprl,vaidman4}.  On the basis of the foregoing argument, that conclusion must be invalid because it is impossible to put a quantum system into a measurement-ready condition for two components of spin at any single time. It is preferable to have a theory that avoids the assignment of definite, dispersion-free values to non-commuting observables because those values cannot in general satisfy the same functional relationships that the observables themselves obey.

\subsection{3-box problem}

In this {\it gedanken} experiment \cite{aaaprl}, the quantum system in the RW and the CFW is described by a three dimensional  Hilbert space  spanned by the rays $|x_i \rangle $, $i=1,2,3$ which can be thought of as representing occupancy of three different boxes. The quantum system is prepared (preselected) at $t_a$ in the state
\begin{equation}
|a \rangle = \frac{1}{\sqrt{2}}(|x_1 \rangle + |x_2 \rangle ) 
\end{equation}
and measured (postselected) at $t_b$ to be in the state
\begin{equation}
|b \rangle = \frac{1}{\sqrt{2}}(|x_2 \rangle + |x_3 \rangle ). 
\end{equation}

The 3-box problem involves the consideration of two observables in the period $t_a < t < t_b$: observable $X$ with eigenstates $|x_i \rangle$, $i=1,2,3$ and observable $Q$ with eigenstates
\begin{subequations}
\begin{eqnarray}
|q_1 \rangle & = & \frac{1}{\sqrt{2}}(|x_1 \rangle + |x_3 \rangle ) \\
|q_2 \rangle & = & |x_2 \rangle  \\
|q_3 \rangle & = & \frac{1}{\sqrt{2}}(|x_1 \rangle - |x_3 \rangle ).\\
\end{eqnarray}
\end{subequations}

\begin{figure}
\includegraphics[scale=0.4]{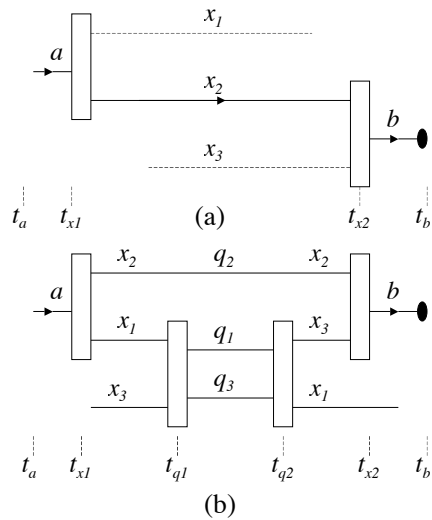}
\caption{(a) A possible sequence of physical interactions that could be used for counterfactual reasoning for observable $X$ in the 3-box problem. Given the preselection for state $|a \rangle$  and the postselection for state $|b \rangle$ as shown, the outcome $|x_2 \rangle$ for the measurement of $X$ in the CFW is certain and therefore property $x_2$ is an element of reality in the RW. (b) A possible sequence of physical interactions that could be used for counterfactual reasoning for observable $X$ and/or of observable $Q$ with the common eigenstate $|x_2 \rangle \equiv |q_2 \rangle $ . The transformation at $t_{q1}$ is the inverse of that at $t_{q2}$. For a measurement of $X$ without $Q$ in the CFW, the outcome $|x_2 \rangle$ or equivalently $|q_2 \rangle$ remains certain but for a measurement of $Q$ with or without $X$ in the CFW, the outcome $|q_2 \rangle$ or equivalently $|x_2 \rangle$ is not certain.}
\label{fig2}
\end{figure}

\noindent
\underline{\it Counterfactual reasoning based on observable $X$.} \\

\noindent
One possible sequence of physical interactions that could be used for the counterfactual reasoning for observable $X$ is shown in Fig. 2(a).
For counterfactual reasoning based on Eq. (3), the following properties hold in the RW at the indicated times
\begin{subequations}
\begin{eqnarray}
\text{Pr}^{\text{RW}}[a|a,A,b] & = & 1 \;\;\; t_a < t < t_{x1} \\
\text{Pr}^{\text{RW}}[x_1|a,X,b] & = & 0 \;\;\; t_{x1} < t < t_{x2}\\
\text{Pr}^{\text{RW}}[x_2|a,X,b] & = & 1 \;\;\; t_{x1} < t < t_{x2}\\
\text{Pr}^{\text{RW}}[x_3|a,X,b] & = & 0 \;\;\; t_{x1} < t < t_{x2}\\
\text{Pr}^{\text{RW}}[b|a,B,b] & = & 1\;\;\; t_{x2} < t < t_b. 
\end{eqnarray}
\end{subequations}
It follows that box or path $x_2$ is an element of reality in the above sense for the given preselection and postselection conditions during the period $t_{x1} < t < t_{x2}$.

If the counterfactual reasoning is based instead on Eq.~(2), the results in Eq.~(8) apply but without the time constraints. Using that approach as the basis of the counterfactual reasoning, Albert, Aharonov and D'Amato \cite{aaaprl} concluded that the three non-commuting observables $A$, $X$ and $B$ can be simultaneously well-defined in the interval $t_a < t < t_b$. Clearly, that conclusion does not follow if the counterfactual reasoning is based on Eq.~(3) because none of the results for $A$, $X$ and $B$ apply at the same time. \\

\noindent
\underline{\it Counterfactual reasoning based on observables $X$ and $Q$}.\\

\noindent
It also follows from the ABL rule in Eq. (1) that
\begin{subequations}
\begin{eqnarray}
\text{Pr}^{\text{ABL}}[x_1|a,X,b] = \text{Pr}^{\text{ABL}}[x_3|a,X,b]  = 0 \\
\text{Pr}^{\text{ABL}}[q_1|a,Q,b] \neq 0.
\end{eqnarray}
\end{subequations}
It follows from counterfactual reasoning on the basis of Eq. (2), i.e. without the constraint that the quantum system in the RW be in a measurement-ready condition, that
\begin{subequations}
\begin{eqnarray}
\text{Pr}^{\text{RW}}[x_1|a,-,b] & = & \text{Pr}^{\text{RW}}[x_3|a,-,b]  = 0 \\
\text{Pr}^{\text{RW}}[q_1|a,-,b] & \neq & 0.
\end{eqnarray}
\end{subequations}

As noted in Ref.~\onlinecite{aaaprl}, from Eqs. (7), the ray $|q_1 \rangle $ lies in a subspace spanned by the rays $|x_1 \rangle $ and $| x_3 \rangle $. Eq. (10) shows that it follows from the counterfactual reasoning of TSQT that the probability of a property corresponding to a ray (here $|q_1 \rangle $ ) in a subspace $\mathcal{S}$ can be non-zero while the probability of properties corresponding to rays which span $\mathcal{S}$ ($|x_1 \rangle $ and $|x_3 \rangle $ ) are zero. As a consequence, it is argued in Ref.~\onlinecite{aaaprl} that an assumption made in the proofs of the theorems of Kochen and Specker \cite{k-s} and Gleason \cite{gleason} (and also other ``no-go" theorems \cite{bellrevmodphys}) is not satisfied for quantum systems within the interval between two measurements.

On the basis of the present approach, in order to conduct counterfactual reasoning based on observables $X$ and $Q$, it is necessary to put the quantum system in a measurement-ready condition for both observables. A sequence of interactions like that shown in Fig.~2(b) is one of the possibilities which are suitable for counterfactual reasoning involving both observable $X$ and observable $Q$. It is clear immediately that, except for the common property $x_2 \equiv c_2$, the properties pertaining to $X$ and $Q$ do not apply to the quantum system at the same time. It would seem for that reason alone, results like those in Eq. (10a) could not be combined with that in Eq. (10b) to draw counterfactual conclusions. 

There is an additional reason that the conclusions of Ref.~\onlinecite{aaaprl} do not apply. Firstly it is important to note that the transformation at $t_{q2}$ is the inverse of the transformation at $t_{q1}$, so if the quantum system were on path $x_1$ at $t < t_{q1}$ it must be on path $x_1$ at $t > t_{q2}$ if no measurement of $Q$, i.e. of paths $q_2$ or $q_3$, is performed. Thus even though the quantum system in the RW has been placed in a measurement-ready condition for both $X$ and $Q$, if the counterfactual reasoning is based only a measurement of $X$ in the ABL rule applying in the CFW, the conclusions will be the same as when the quantum system in the RW had not been put in a measurement-ready condition for $Q$, i.e. $x_2$ remains an element of reality in that case.

On the other hand if the counterfactual reasoning is based on a measurement of $X$ and $Q$, the ABL rule for more than 2 intermediate measurements \cite{abl} must be applied in the CFW. The measurement of $Q$ in the CFW means that the quantum system incoming on path $x_1$ for $t < t_{q1}$ may emerge at $x_1$ or $x_3$ for $t > t_{q2}$. It is then not the case that $x_2$ is an element of reality because there is a non-zero probability from the ABL rule that the quantum system will follow the sequence $a \rightarrow x_1 \rightarrow q_1 (\mbox{ or } q_3) \rightarrow x_3 \rightarrow b$ and therefore the probability of the sequence $a \rightarrow x_2 (\equiv q_2) \rightarrow b$ is not unity. Consequently, the conditions in Eq. (9) do not apply because the probabilities of $x_1$ and $x_3$ are non-zero when the probability of $q_1$ is non-zero. 

Thus if the counterfactual reasoning is conducted on the basis of the definitions proposed here, the argument about the 3-box problem in Ref. 2 referred to above and its consequences for the well-known theorems do not follow.   

\section{Conclusion}

Counterfactual reasoning based on TSQT has been a productive concept. It has been suggested here that the counterfactual reasoning should be subject to a qualification.  The qualification is motivated by the fact that the ABL formalism involves an intermediate measurement or sequence of intermediate measurements between preparation and final measurement. Consequently counterfactual reasoning based on the ABL formalism should only be applied to a quantum system which has been subjected to physical interactions so that it is in ``measurement-ready" condition for the measurement required for the ABL rule to apply in the CFW. One of the characteristics which is unique to a ``measurement-ready" condition is that it could be ascertained, on some repetitions of the experiment, that the quantum system was in any one of the states that could be the outcome of the measurement by monitoring the {\it other} possible outcome states, i.e. without in any way disturbing the state the quantum system was thereby found to be in. Of course, that condition is met only on those repetitions when the quantum system is {\it not} found in the other possible outcome states. It is important to re-iterate that being put in a ``measurement-ready" condition is a reversible process and the quantum system is not coupled to a measuring device in the process.   

The qualification means that counterfactual reasoning in TSQT, or which is otherwise based on the ABL formalism, should satisfy Definition A, and Definition B where appropriate, and be calculated according to Eq. (3) rather than Eq. (2). Counterfactual reasoning based on Eq. (3) rather than Eq. (2) leads to more physically reasonable results. It eliminates the ascription of simultaneous dispersion-free values to non-commuting observables and avoids conflict with a fundamental assumption in the derivation of well-known theorems. Furthermore definitions A and B above make the RW and the CFW ``closer" to each other, a desirable property for counterfactual reasoning

The qualification of being ``measurement-ready" in Definitions A and B has not arisen in counterfactual reasoning before because it involves considerations unique to quantum systems. A measurement of a quantum system involves the quantum system being subjected to suitable physical interactions, after which a measurement of a non-commuting observable cannot take place unless the quantum system is subject to different physical interactions. This necessity of subjecting the quantum system to suitable, mutually exclusive physical interactions needs to be reflected in the counterfactual reasoning. Therefore counterfactual reasoning based on {\it gedanken} experiments on a quantum system is only valid when the quantum system has been subjected to the physical interactions necessary for the {\it gedanken} experiment to be carried out in principle. The same qualifications to counterfactual reasoning are not necessary in the case of classical physics because a classical system is in a measurement-ready condition for any measurement at all times.
  
\section{acknowledgments}

I thank R. Kastner and L. Vaidman for their comments on a draft.

\end{document}